\documentstyle[12pt]{article}
\begin{document}
\begin{titlepage}
\begin{flushright}
TIFR/TH/95-29\\
June 1995
\end{flushright}
\begin{center}
\thispagestyle{empty}
{\Large \bf The Computational Complexity of Symbolic Dynamics \\
at the Onset of Chaos}
\end{center}
\bigskip
\begin{center}
 Porus Lakdawala\footnote{E-mail: {\tt porus@theory.tifr.res.in}}\\
Tata Institute of Fundamental Research\\
Homi Bhabha Road, Bombay-400005,INDIA.
\end{center}
%\date{June 1995.}
%\maketitle
\begin{abstract}
  In a variety of studies of dynamical systems, the edge of order and
  chaos has been singled out as a region of complexity. It was
  suggested by Wolfram, on the basis of qualitative behaviour of
  cellular automata, that the computational basis for modeling this
  region is the Universal Turing Machine. In this paper, following a
  suggestion of Crutchfield, we try to show that the Turing machine
  model may often be too powerful as a computational model to describe
  the boundary of order and chaos.  In particular we study the region
  of the first accumulation of period doubling in unimodal and bimodal
  maps of the interval, from the point of view of language theory. We
  show that in relation to the ``extended'' Chomsky hierarchy, the
  relevant computational model in the unimodal case is the nested
  stack automaton or the related indexed languages, while the
  bimodal case is modeled by the linear bounded automaton or the
  related context-sensitive languages.
\end{abstract}
\vskip 1in
\smallskip
\end{titlepage}

\renewcommand{\thefootnote}{\arabic{footnote}}
\setcounter{footnote}{0}

\section*{\protect\normalsize\bf 1. Introduction}

The complex systems that we often observe, both in nature and
otherwise, are characteristically poised in a delicate balance between
the dullness of order and the randomness of disorder. In recent years
a lot of effort has been expended in obtaining quantitative measures
which would provide suitable definitions for this complexity.

In a study of the qualitative behaviour of cellular automata, Wolfram
\cite{wolf} noticed that there was a class of automata, which he
called Class 4, whose members displayed complex dynamical behaviour.
In particular, they seemed to exhibit transients which were
arbitrarily long-lived.  Wolfram suggested that this behaviour was
reminiscent of the undecidability which characterised the halting
problem for Turing machines. He conjectured that Class 4 automata
then might have a universal Turing machine embedded within them.

The model of the Turing machine has served, in the past, to provide a
deeper understanding of phenomena which had been investigated
through other means. The most notable amongst these is the development
of the idea of algorithmic information which is the computational
counterpart of the traditional Shannon entropy. These ideas have
served to provide a quantitative basis for the qualitative notion of
randomness.  Conceived originally through the efforts
of a number of people, it has shed light on some deep issues related
to the foundations of mathematics, mainly through the work of Chaitin
\cite{chaitin}.

In \cite{avinash} we tried to show that cellular automata, with
Turing machines embedded within them, may not always display complex
behaviour. In our present work we try to demonstrate that the model of
the Turing machine may at times be too powerful to describe the
computational complexity at the edge of order and chaos.
Crutchfield \cite{crutchfield} has argued that instead of looking
only at the Turing machine, it might
be wiser to look at the entire hierarchy of machines that computation
theory provides us with. Classical computation theory provides a
beautiful framework in which models of machines or automata are
related to the languages they recognise, which are in turn, generated
by their respective grammars. This hierarchy is known as the Chomsky
hierarchy. The Turing machines form the top of this hierarchy.

The dynamical systems we study are, the rather well understood,
iterated maps of the interval. We investigate the onset of
chaos which is exhibited by the first accumulation of period doubling
for the case of the unimodal and bimodal families of maps. The
symbolic dynamics and the kneading theory for these cases are well
known \cite{collet,mackay}. We demonstrate that the language generated
by the kneading sequences in these two cases can be recognised by machines
which lie lower down in the Chomsky hierarchy.

\section*{\protect\normalsize\bf 2. The symbolic dynamics of
Maps of the interval}

In this section we set up the notation used by Mackay and Tresser
\cite{mackay} to describe the onset of chaos at the first
accumulation of period doubling.

\subsection*{\protect\normalsize 2.1. The unimodal case}

A unimodal map is, by definition, a continuous map $f$ of the interval
$I=[0,1]$ into itself, which possess a single turning point $c$.
Normally $f$ is chosen to so that it monotonically increases in the
interval $[0,c)$ and decreases monotonically in the interval $(c,1]$.
With every point in $I$ we associate a symbol $L$ or $R$ depending on
whether it lies on the left or on the right of $c$. $c$ is identified
with the symbol $C$.  In this way the orbit of any point in $I$ under
iteration by the map $f$ can be associated with a sequence
\footnote{As is done conventionally, an eventually periodic sequence
of symbols will be described as a finite sequences by enclosing the
periodic part in brackets with the symbol ``$\infty$'' at the end as in
$\cdots(\cdots)^\infty$}of symbols
from the set $\Lambda={L,C,R}$.  The kneading sequence which is defined to
be the symbol-sequence corresponding to the orbit of the point $f(c)$,
is of particular importance in the symbolic dynamics. It has be shown,
in the pioneering work of Milnor and Thurston \cite{milnor}, that the
kneading sequence controls the possible symbol-sequences that can occur
for a given map and that it determines important ergodic properties
of the map, like the topological entropy.

In order that we can define the symbol-sequences of interest to us, we
need mention here only a single detail from symbolic dynamics.
Symbolic sequences can be ordered
in a way, such that they respect the ordering of the interval $I$ {\it i.e.}
if $x,y \in I$ have symbol-sequences $s(x)$ and $s(y)$ respectively, then
\[  x \leq y \mbox{ iff } s(x) \leq s(y). \]
\noindent This ordering is defined as follows:

First we define an ordering on the symbols as $ L < C < R $.
Now, if $A = Xa\ldots$ and $B = Xb\ldots$ are two sequences, for which
X is the common prefix-sequence and $a,b \in \{L,C,R\},~
a \neq b$ then
\begin{displaymath}
\begin{array}{ll}
A < B &\mbox{if $a < b$ and $X$ contains an even number of $R$'s or}  \\
      &\mbox{ \,\,\,\,$a > b$ and $X$ contains an odd number of $R$'s.}  \\
A > B &\mbox{otherwise}
\end{array}
\end{displaymath}

It is well known that families of unimodal maps (like the well-known
logistic family) exhibit the route to chaos through period-doubling
transitions. The renormalisation group theory pioneered by Feigenbaum
\cite{feigenbaum} provides a beautiful description of this phenomenon.
For our purposes it suffices to describe the kneading sequences that
arise through successive applications of the renormalisation group
transformation. The limit sequence then can be thought of as the
symbolic description of the onset of chaos, obtained through an
accumulation of period-doubling transformations. The algorithm for
generating these sequences is as follows:
\begin{eqnarray*}
X^{(0)} & = & \emptyset  \\
X^{(n+1)} & = & X^{(n)}\bar{U}^{(n)}X^{(n)}.
\end{eqnarray*}
where the $U^{(n)}$ and $\bar{U}^{(n)}$ are chosen from $\{L,R\}$
so that
\begin{displaymath}
X^{(n)}U^{(n)} < X^{(n)}C < X^{(n)}\bar{U}^{(n)}.
\end{displaymath}
We mention that this algorithm generates the kneading sequences
$K_n = (X^{(n)}C)^\infty,n \geq 0$
of maps, having a superstable periodic orbit of period $2^{n}$.
Moreover, the set of periods of periodic points of the map
with the kneading sequence $K_n$, is given by $\{1,2,4,\ldots,2^n\}$.
Each of these maps have a finite, stable attracting set and
hence display ordered dynamics.

We also define another set of kneading sequences which would be useful
for the discussion
\begin{displaymath}
K'_n =
(X^{(n)}\bar{U}^{(n)}X^{(n)}U^{(n)}X^{(n)}\bar{U}^{(n)})^\infty,~~n \geq 0
\end{displaymath}
The sequence $K'_n$ represents a map which contains an orbit of
period $3\cdot 2^n$. These maps have positive topological entropy
and can, in this sense, be considered as displaying chaotic dynamical
behaviour.

The sequence $K_\infty = K'_\infty$, is the kneading sequence of
the map at the onset of chaos. The set of periods, of the
periodic points of this map is given by  $\{1,2,4,\ldots,2^n,\ldots \}$.
The topological entropy in this case is zero.

For the details of these results, we refer the reader to the
literature \cite{collet,mackay}.

\subsection*{\protect\normalsize 2.2. The bimodal case}

We now describe the accumulation of period-doubling in bimodal maps
through their symbolic dynamics. Once again we refer the interested
reader to \cite{mackay} for the details.

A bimodal map of the interval $I$, is a map $f$ from $I$ into itself
with two turning points, $c$ and $k$. In what follows we consider the
case of the $+-+$ maps, {\it i.e.} maps which are monotonic increasing
on $(0,c)$ and $(k,1)$, monotonic decreasing on $(c,k)$.  Each point
in $I$ is now associated with a symbol from the set $\{L,R,B\}$
depending on whether it belongs to the interval $(0,c)$, $(c,k)$ or
$(k,1)$ respectively. The points $c$ and $k$ are assigned the symbols $C,K$
respectively. The symbol-sequence corresponding to the orbit of a
point, consists of a string of symbols from the set $\Sigma =
\{L,C,R,K,B\}$.
In this case, the kneading data of a map corresponds to a pair of
kneading sequences, which correspond to the symbol-sequences of
the pair of points $(f(c),f(k))$. As mentioned before, the kneading
data uniquely determine some important properties of the map.

The definition of the ordering on symbol-sequences for the unimodal case
from the previous section,
can be retained verbatim for the bimodal case, if we define the
ordering on the elementary symbols as $L < C < R < K < B$. The
parity of common prefix of two symbol strings is still determined
by the number of $R$'s in it. This is due to the fact that $R$
corresponds to the symbol for that portion of the interval $I$, on which
$f$ is monotonic decreasing. Thus defined, the ordering on symbol-
sequences respects the ordering on the interval as before.

We now give the description of the onset of chaos, corresponding to the
first accumulation of period-doubling for bimodal maps. We define operations
$l$ and $r$ on pairs $(X,Y)$ of finite (possibly empty) sequences of
$\{L,R,B\}$ by
\begin{equation}
(X,Y)\, l = (X,Y\bar{V}XUY), ~~~(X,Y)\, r = (X\bar{U}YVX,Y)
\label{basicops}
\end{equation}
where $U,\bar{U}$ are chosen from $\{R,B\}$ so that $XU < XK <
X\bar{U}$ and $V,\bar{V}$ are chosen from $\{L,R\}$ so that
$Y\bar{V} < YC < YV$.

Given a finite sequence $s$ of $l$'s and $r$'s, we define
\begin{equation}
(X_{s},Y_{s}) = (\emptyset,\emptyset)s
\label{treepath}
\end{equation}
and write $U_{s},\bar{U}_{s},V_{s},\bar{V}_{s}$ for
the $U,\bar{U},V,\bar{V}$ corresponding to $X_{s},Y_{s}$.

Pairs $(P,Q)$ of sequences (possibly infinite) over the symbol-set
$\Sigma$ can be put into a one-to-one correspondence with points in the
unit square (see Figure 1({\it i})), which we denote by $G$.
For $(P_{0},Q_{0}) \in G$,
we define a {\it wedge} (see Figure 1({\it ii}))
\begin{displaymath}
\omega (P_{0},Q_{0}) = \{(P,Q) \in G:P \geq P_{0},Q \leq Q_{0}\}.
\end{displaymath}
Then define
\begin{displaymath}
R_{n} = \bigcup_{\| s \| = n}
         \omega ((X_{s}U_{s}Y_{s}C)^\infty,(Y_{s}V_{s}X_{s}K)^\infty)
                                ~~~{\rm for}~~~ n \geq 0
\end{displaymath}
where $\| s \|$ is the length of $s$.

All maps on the boundaries of the wedges defining $R_{n}$ have a singly
superstable, period $2^{n+1}$ orbit passing through $C$ or $K$.

We also define regions $S_{n}, n=1,2,3 \ldots$, such
that $S_{n}$ contains maps which have certain period $3\cdot2^{n-1}$ orbits.
The definitions of these $S_{n}$ is cumbersome. We reproduce it here
for completeness.
\begin{displaymath}
\begin{array}{lc}
S_{1} = \omega ((RLR)^\infty,(LRR)^\infty) \bigcup
                         \omega ((BRR)^\infty,(RBR)^\infty)
 &~~~~~~~~~~~~~~~~~~~~~~~~~~~~~~~~~~~~~~~~~~~~~~~
\end{array}
\end{displaymath}
while for $n \geq 2$,
\begin{eqnarray*}
S_{n} = \bigcup_{\| s \| = n-2} &
\!\!\!\left\{~\omega
\left(\left((XUY\bar{V}XUYVXUY\bar{V})_{s}\right)^\infty\right.,
         \left((Y\bar{V}XUYVXUY\bar{V}XU)_{s}\right)^\infty \right) &  \\
&~~~~ \bigcup
\ \omega \left(\left((X\bar{U}Y\bar{V}XUY\bar{V}XUY\bar{V})_{s}\right)^\infty,
      \left((Y\bar{V}XUY\bar{V}X\bar{U}Y\bar{V}XU)_{s}\right)^\infty\right) &\\
&~~~~ \bigcup
 \ \omega \left(\left((X\bar{U}YVX\bar{U}Y\bar{V}X\bar{U}YV)_{s}\right)^\infty,
      \left((Y\bar{V}X\bar{U}YVX\bar{U}YVX\bar{U})_{s}\right)^\infty\right) &
\\
&~~~~~~ \bigcup
  \   \omega \left(\left((X\bar{U}YVXUYVX\bar{U}YV)_{s}\right)^\infty,
            \left((YVX\bar{U}YVXUYVX\bar{U})_{s})^\infty \right)\right\}  &
\end{eqnarray*}
where $(ABC\ldots)_{s} \equiv (A_{s}B_{s}C_{s}\ldots)$.

The regions $\bar{R}_1$ and $S_1$ are shown in Figure 1({\it iii}).

With these definitions, following \cite{mackay}, we define
subsets of $G$, which will serve as the regions of order and chaos.
\begin{eqnarray*}
\bar{R}_{\infty} & = &  \bigcup_{n} \bar{R}_{n}
                                ~~ = ~~ \bigcup_{n} (G\backslash R_{n}) \\
S_{\infty} & = & \bigcup_{n} S_{n} \\
D~~ & = &  G\backslash (\bar{R}_{\infty} \bigcup S_{\infty})
\end{eqnarray*}

Theorem 1 of \cite{mackay} tells us, that the set of periods of a map
contained in the region $\bar{R}_{\infty}$
is given by $\{1,2,4,\ldots ,2^n\}$ for some $n$. Moreover all such
maps are contained within it. Thus maps within this region exhibit
ordered dynamics (and have zero topological entropy).

On the other hand, a map contained in the region $S_{\infty}$
has at least one periodic orbit with a period, which is not a
power of $2$. Moreover all such maps belong to this region and have
positive topological entropy. Hence, we consider $S_\infty$ as
the region of chaos.

The set of periods of a map in the boundary $D$ is given by
$\{1,2,4,\ldots,2^n,\ldots\}$. Again, the converse is also true
in this case. We will consider $D$ as the ``boundary'' of order and chaos.
Whether it is a boundary, in a precise topological sense, of the
regions $\bar{R}_{\infty}$ and $S_{\infty}$ is a long-standing conjecture.

\section*{\protect\normalsize\bf 3. The ``extended'' Chomsky hierarchy}

The classical theory of automata and formal languages, has its roots,
on the one hand, in the work of Turing, Church and others who
developed the foundations of the theory of computation and on the other,
in the work of formal linguists, notably Chomsky. In the present section
we present a brief outline of this very large body of work, with an
emphasis on that part which will be relevant for our purpose. For the
details the reader is referred to the classic book by Hopcroft and Ullman
\cite{hopcroft}.

The entities that automata theory deals with are formal descriptions
of devices which can perform ``mechanical'' tasks of varying degrees
of difficulty. These devices usually, have an input tape containing
symbols from an alphabet, a memory unit and an auxiliary tape which is
used to perform storage and output. The dynamics of the automaton is
described by defining the action(s) to be performed, depending on the
symbol being currently read on the input tape, the present memory
state and the current symbol on the storage tape. These actions might
entail a change of the memory state, alteration of the current
symbol on the storage tape and motion along the input tape in
either direction. The Turing machine is probably the best known
example of an automaton. The thesis of Church and Turing states that
it is equivalent to any other system that can perform a task that can
be defined algorithmically.  The central column in Figure 2 shows
various classes of automata, that form a hierarchy of machines in terms
of their computational power.

Formal languages are sets, containing finite strings (called words)
of symbols
from some (finite) alphabet. The most common examples of formal languages are
the programming languages that are used today. These are contained in
the class of formal languages called the context-free languages. One
way of describing formal languages, is through the grammars that
generate them. A grammar contains, not only the symbols which form the
words in the language that it generates (these are called terminal
symbols), but also an auxiliary set of symbols, called the
non-terminal symbols. The ``productions'' of the grammar, form the
rules by which a given string of terminals and non-terminals, can
generate another string of terminals and non-terminals. A special
non-terminal (usually denoted by $S$) is designated the ``start''
symbol. Any string formed
from the terminal symbols only, that can be obtained from the ``start''
symbol, by applying the productions successively, is a valid word in
the language. The columns on the left and right in Figure 2, show
a hierarchy of the classes of languages and their generative
grammars. It is a beautiful fact of computation theory that the
languages generated by the different grammars, are also ``recognised''
by the various classes of automata shown in Figure 2.

This hierarchy of languages, grammars and machines goes under the name
of the Chomsky hierarchy\footnote{Conventionally the Chomsky hierarchy
does not contain the indexed and the stacked languages and the automata
that recognise them. We call this the ``extended'' Chomsky hierarchy.}.

A very natural problem that arises in computation theory, is to determine
the class in which a given language belongs. For this purpose, the
theory provides a powerful set of results which are called intercalation
theorems or sometimes pumping lemmas. These are necessary conditions
for a language to belong to a particular class. The usual application
of a pumping lemma is to exclude a given language from a certain class.
Most commonly, the form of a pumping lemma is as follows: Given any
word belonging to a language (in a certain class), it is possible to
find a sub-word(s), of this word, of length not greater than $k$ (an integer
that depends only on the language and not the word) such that all the
words obtained by ``pumping'' ({\it i.e.} adding another copy of the sub-word
at its position in the given word) the sub-word finitely often, would
all belong to the language. In the next section we use the pumping
lemmas for the one-way, non-deterministic stacked languages and the
indexed languages.

\section*{\protect\normalsize\bf 4. The results and proofs}

In this section we present the main results of this paper and their
proofs. We first define formal languages corresponding to the
onset of chaos in the unimodal and bimodal cases and then obtain
the classes in the Chomsky hierarchy where these languages occur.
The key ingredient in each proof is a relevant pumping
lemma. As the statements of these lemmas are rather involved,
we will use the lemma and refer the reader to the
literature for its precise statement.

\subsection*{\protect\normalsize 4.1. The unimodal case}

In the notation of section 2.1, we introduce a sequence of languages
${\cal A}_{n} = \{ K_{r} : 0 \leq r \leq n\}$. The words in these languages are
formed from the alphabet $\Lambda' = \{ L, C, R\} \bigcup \{(,)^\infty\}$.
Moreover for each $n$,
${\cal A}_{n}$ is finite and is contained in ${\cal A}_{n+1}$. Thus each
${\cal A}_{n}$ is a
regular language. The limit language ${\cal A}_{\infty}$ can be thought
of as a symbolic description of the edge of order and
chaos corresponding to the sequence
of period-doublings of unimodal maps. Our concern here is to
describe the place, that this language has in the ``extended'' Chomsky
hierarchy. This is summarised by the following theorem,

\vskip 0.25in
\noindent {\bf Theorem 4.1}
\begin{enumerate}
\item[{\it (i)}] ${\cal A}_{\infty}$ is not a one-way, non-deterministic stack
language.
\item[{\it (ii)}] ${\cal A}_{\infty}$ is an indexed language.
\end{enumerate}

\noindent Proof:

{\it (i)} We use the pumping lemma for one-way, non-deterministic stack
languages, which can be found in \cite{ogden}. Suppose ${\cal A}_{\infty}$ is
a one-way, non-deterministic stacked language. Consider the integer
$k$ in Theorem 1 (henceforth referred to as OG1) of \cite{ogden}. Let
${\xi}_{0}$ be a word in ${\cal A}_{\infty}$ of length greater than $k$. The
pumping lemma then guarantees the existence of a string of words
${\xi}_{i}, i = 1,2,3 \ldots$ obtained by intercalating strings within
${\xi}_{0}$ as described in OG1. If $\|\xi\|$ denotes the length of the
word $\xi$, we have from OG1, for all $i > 0$,
\begin{eqnarray}
\|{\xi}_{i}\| - \|{\xi}_{i-1}\| =& \|{\rho}_{i}\| + \|{\sigma}_{i}\| +
\|{\tau}_{i}\| -
                              \|{\sigma}_{i-1}\|       \nonumber  \\
                            =& a + ib
\label{xiequation}
\end{eqnarray}
where $a$ is the sum of the lengths of all the
${\alpha}_{j}$'s, ${\gamma}_{j}$'s,
${\delta}_{j}$'s, ${\chi}_{j}$'s, ${\psi}_{j}$'s and $b$ is the sum of all
${\beta}_{j}$'s that appear in the definitions of
${\rho}_{i},{\sigma}_{i}$ and ${\tau}_{i}$.

(\ref{xiequation}) is easily solved recursively, to give
\begin{equation}
P(i) \equiv \|{\xi}_{i}\| = \|{\xi}_{0}\| + ai + bi(i-1)/2
\end{equation}
Now observe that for a word $\lambda$ in
${\cal A}_{\infty}$, $\|\lambda\| = 2^{m}+2$ for some $m \geq 0$.
Since each ${\xi}_{i}$ is in ${\cal A}_{\infty}$ we have
\begin{equation}
 P(i) = 2^{m} + 2
\label{P}
\end{equation}
for every $i \geq 0$, for some $m \in {\bf N}$.
That this is impossible is most easily seen as follows:

\noindent $P'(x) \equiv P(x) - 2$ is a polynomial taking integer values
on the set of integers. Hence
the set $\{P'(n) : n \in {\bf N} \}$ has an infinite number of prime divisors
(this is rather easy to prove). But from (\ref{P}) $P'(x)$ is contained
in $\{ 2^{n} : n \in {\bf N} \}$.
This leads to a contradiction and completes the proof.

\vskip 0.25in

{\it (ii)} The simplest proof of the fact that
${\cal A}_{\infty}$ in an indexed language is probably obtained by observing
that ${\cal A}_{\infty}$ is a $0L$ language (see \cite{hopcroft} pg. 390ff,
for the definition) and hence is also an indexed language. We will, however,
explicitly define the indexed grammar which generates ${\cal A}_{\infty}$.

Consider the indexed grammar defined as:

\noindent The non-terminal symbols are $\{ S,T,X,U,\bar{U} \}$, with
$S$ as the start symbol.
\noindent The terminal symbols are $\{ L,R,C,(,)^\infty \}$.
\noindent The index-productions are (here $\varepsilon$ is the empty string)
\begin{eqnarray*}
Xf & \longrightarrow & X\bar{U}X  \\
Uf & \longrightarrow & \bar{U} \\
\bar{U}f &  \longrightarrow & U  \\
Xg & \longrightarrow & \varepsilon    \\
Ug & \longrightarrow & L  \\
\bar{U}g &  \longrightarrow & R  \\
\end{eqnarray*}
\noindent The productions are
\begin{eqnarray*}
S & \longrightarrow & (C)^\infty  \\
S & \longrightarrow & (TgC)^\infty  \\
T & \longrightarrow & Tf   \\
T & \longrightarrow & X\bar{U}X   \\
\end{eqnarray*}

The derivation of the words in ${\cal A}_\infty$ using the indexed
grammar proceeds as follows:

As is conventional, $A \Rightarrow B$ means that $B$ is derived from
$A$ using a single production (or index production), while
$A \stackrel{*}{\Rightarrow} B$ means that there exist
$C_1,C_2,\ldots,C_n$ such that
$A \Rightarrow C_1  \Rightarrow C_2 \Rightarrow \cdots \Rightarrow C_n
\Rightarrow B$. Here $A,B,C_i$ are words formed from the terminal and
non-terminal symbols
\begin{displaymath}
\begin{array}{ccccccccc}
S & \Rightarrow   & (C)^\infty    & & & & & & \\
S & \Rightarrow   & (Tg)^\infty  & \Rightarrow & (Xg\bar{U}gXgC)^\infty &
\stackrel{*}{\Rightarrow} & (RC)^\infty & & \\
  & & & & & & & & \\
  & & \Downarrow & & & & & & \\
  & & & & & & & & \\
  & & (TfgC)^\infty & \Rightarrow & (Xfg\bar{U}fgXfgC)^\infty
  & & & & \\
  & & \Downarrow & & & & & & \\
  & & \vdots & & \Downarrow * & & & & \\
  & & \Downarrow & & & & & & \\
  & & \vdots & & (Xg\bar{U}gXgUgXg\bar{U}gXgC)^\infty
  & \stackrel{*}{\Rightarrow} & (RLRC)^\infty & &  \\
  & & \Downarrow & & & & & & \\
  & & \vdots & & & & & & \\
  & & (T\underbrace{ff\cdots f}_{\mbox{n times}}gC)^\infty
  & \cdots & & & & & \\
  & & \vdots & & & & & & \\
\end{array}
\end{displaymath}

\subsection*{\protect\normalsize 4.2. The bimodal case}

We define the language which would describe the first
accumulation of period-doubling for the bimodal maps.
In the terminology of section 3b, we describe each wedge $\omega(P,Q)$
by the word $\langle P:Q\rangle$. The region $R_{n}$ is obtained as a union
of wedges, each of which corresponds to a unique finite string $s$ of
$l$'s and $r$'s of length $n$.

To describe the word corresponding to $R_{n}$ we first introduce the
lexicographical ordering on strings $s$, induced by the ordering $l <
r$ on the symbols of $s$ {\it i.e.} if $s_{1} = xa\ldots$ and $s_{2} =
xb\ldots$ are two strings on $\{l,r\}$, with a common prefix-sequence
$x$ and
$a,b \in \{l,r\}, a \neq b$, then $s_{1} < s_{2}$ iff $a < b$, else
$s_{1} > s_{2}$. Note that this ordering is different from the
ordering on words we had defined in section 3.1.

\vskip 0.15in
\noindent{\bf Remark (1)}:
The words $s$ can be generated as the nodes of a binary
tree. The lexicographical ordering is then the conventional ordering
of nodes from left to right of a binary tree (see Figure 3).
We will refer to this tree as ${\cal T}_{s}$.
\vskip 0.15in
\noindent With this we now define the word corresponding to $R_{n}$
as follows:
\begin{enumerate}
\item[(a)] Arrange the wedges occurring in the definition of $R_{n}$, in
increasing lexicographical order of the strings $s$ that each wedge
corresponds to.

\item[(b)] Concatenate the words corresponding to each wedge in
the same order to form the word corresponding to $R_{n}$.
\end{enumerate}
In exactly the same fashion we can obtain the words which describe
$S_{n}$. Let $w(R_{n})$ and $w(S_{n})$ be the words which
describe $R_{n}$ and $S_{n}$ respectively.

\vskip 0.15in
\noindent{\bf Remark (2)}:
 The length of the word representing a wedge in the definition
of $R_{n}$ is of length $2^{n+2} + 7$. This follows from the fact that
if $s$ has length $n$, then the lengths of $X_{s}$ and $Y_{s}$ add up
to $2^{n+1}-2$. Similarly the length of a word that describes a wedge
in $S_{n}$ is $3\cdot 2^{n} + 7$. In particular, note that the number
of symbols that occur between two consecutive occurrences of the symbol
``$\langle$'', in the definitions of $R_n$ and $S_n$, is a (non-constant)
function of $n$.

\vskip 0.15in
We define a sequence of languages ${\cal B}_{i}, i = 1,2 \ldots$
on the alphabet
\[\Sigma^\prime =
\Sigma ~ \bigcup ~\{\langle,\ \rangle,\ :,\ ;,\ (,\ )^\infty\}\]
as follows:
\begin{displaymath}
{\cal B}_{i} = \{ w(R_{j}) : j = 1,2\ldots i\}.
\end{displaymath}
This language can be thought of as describing the region
$\bigcup_{j=1}^{i} {\bar{R}}_{i}$.

As an example, we construct the language
${\cal B}_{1}$.
\begin{displaymath}
\begin{array}{l}
X_{\emptyset} = \emptyset,~Y_{\emptyset} = \emptyset,~U_{\emptyset} = R,
 ~\bar{U}_{\emptyset} = B,~V_{\emptyset} = R,~\bar{V}_{\emptyset} = L \\
X_{l}=\emptyset,~Y_{l}=LR,~U_{l}=R,~\bar{U}_{l}=B,~V_{l}=L,~\bar{V}_{l}=R  \\
X_{r}=BR,~Y_{r}=\emptyset,~U_{r}=B,~\bar{U}_{r}=R,~V_{r}=R,~\bar{V}_{r}=L  \\
\\
w(R_{1})=\langle(RLRC)^\infty:(LRLK)^\infty\rangle\langle (BRBC)^\infty:
(RBRK)^\infty\rangle   \\
{\cal B}_{1}= \{\langle(RLRC)^\infty:(LRLK)^\infty\rangle\langle (BRBC)^\infty:
(RBRK)^\infty\rangle \}
\end{array}
\end{displaymath}

The limit language ${\cal B}_\infty$ can be thought of as
describing the ``boundary'' $D$ of order and chaos.
The following theorem characterises
${\cal B}_{\infty}$ with respect to the Chomsky hierarchy.
\vskip 0.25in
\noindent {\bf Theorem 4.2.}
\begin{enumerate}
\item[{\it (i)}] ${\cal B}_{\infty}$ is not an indexed language.
\item[{\it (ii})] ${\cal B}_{\infty}$ is a context-sensitive language.
\end{enumerate}

\noindent Proof:

{\it (i)} The pumping lemma for indexed languages is given in \cite{hayashi}.
Our proof closely follows Theorem 5.3 (henceforth referred to as HA5.3)
of \cite{hayashi}. We
urge the reader to refer to \cite{hayashi} for a description of the
notation that we use in this proof.

Suppose ${\cal B}_{\infty}$ is an indexed language. Choose ``$\langle$''
as a special symbol of $\Sigma^\prime$. With every word of an indexed
language, we can associate a derivation tree (see \cite{hayashi}).
A node, $p$, of a derivation tree, $\gamma$, is said to be a P-node,
if there exist at least two distinct sub-trees under it,
each of which contains at least one node with the label ``$\langle$''.
A pair of nodes $p_1,p_2$ of $\gamma$ are said to be CF-like
if
\begin{enumerate}
\item[(a)] $p_2$ is a descendent of $p_1$
\item[(b)] $p_1$ and $p_2$ have the same labels.
\item[(c)] there exists a P-node $p$, such that $p$ is a descendent
 of $p_1$ and $p_2$ is a descendent of $p$.
\end{enumerate}
If $\gamma$ contains no CF-like pair of nodes, it is said to be
non-CF-like.

We now show that if $\psi$ is a word in ${\cal B}_{\infty}$ of
large enough length, then parts of it can be intercalated in a way
such that the resulting words would not belong to ${\cal B}_{\infty}$.
Choose $\psi$ to be any word in which the number of occurrences of the
symbol ``$\langle$'' is more than $k'$, where $k'$ is an integer
which depends only on ${\cal B}_\infty$
(and is defined in HA5.3). Say $\gamma$ is the derivation tree
of $\psi$. We consider two cases.

\vskip 0.15in

\noindent {\it Case(a}): $\gamma$ is non-CF-like.

In this case, following the proof of HA5.3 we can establish that there
is a decomposition $\gamma = \alpha\cdot\beta\cdot\delta\cdot
\tau\cdot\nu$ such that either $\alpha$ or $\nu$ has at least three
P-nodes. Then by intercalating parts of $\gamma$ we obtain a
sequence of trees $\theta_{n}$, $n = 1,2,3\ldots$ each of which has
$\alpha$ and $\nu$ as its (first and last respectively) components.
Moreover the word $g({\theta}_n)$ of which ${\theta}_n$ is a derivation
tree, belongs to ${\cal B}_{\infty}$ and
$\|g({\theta}_{n})\| < \|g(\theta_{n+1})\|$. This means that there exist
words of increasing length in ${\cal B}_{\infty}$ such that each of them
contains a {\em fixed} sub-word of the form $\langle \cdots \langle$
(because either $\alpha$ or $\nu$ has at least three P-nodes).
This contradicts Remark (2).

\vskip 0.15in

\noindent {\it Case(b)}: $\gamma$ has a CF-like pair of nodes.

In this case the proof of lemma 2.1 of \cite{hayashi} guarantees a
decomposition $\gamma = \alpha\cdot\beta\cdot\delta$ such that $\beta$
contains a P-node. Moreover for each $n = 1,2,3\ldots$,there exists
a sequence of trees
\begin{displaymath}
\gamma_{n} =
\alpha\cdot\underbrace{\beta\cdot\beta\cdots\beta}_{\mbox{n times}}\cdot\delta
\end{displaymath}
such that each $g(\gamma_{n}) \in {\cal B}_{\infty}$
and $\|g(\gamma_{n})\| < \|g(\gamma_{n+1})\|$. This implies that for every
$n \geq 1$, $g(\gamma_{n})$ contains a sub-word of the form
$(\cdots\langle \cdots)^{n}$ (Note that by $\cdots\langle\cdots$ we mean
a {\em fixed} sub-word of that form).
Since the $\gamma_{n}$'s
belong to ${\cal B}_{\infty}$ we have a contradiction with Remark (2).

\vskip 0.25in

{\it (ii)} We prove that ${\cal B}_{\infty}$ is a context-sensitive language by
showing that it in fact belongs to the complexity class DSPACE($n$).
The proof that the class of context-sensitive languages is equivalent
to the complexity class NSPACE($n$) (which contains DSPACE($n$)),
can be found in \cite{hopcroft}.

To show that a language is DSPACE($n$), we need to show that the words
in the language can be recognised by an off-line, multitape Turing
machine such that the size of every tape (including the read-only
tape) is limited to be of the size of the input word (buffered on
either side by end-marker symbols). We will informally describe the
steps in the algorithm required to recognise the input string. It will
be clear from our description that each subroutine in this algorithm
can be implemented on a bounded tape (or sometimes a pair of tapes) of
our machine. The number of subroutines involved in the description
will thus determine the number of tapes of the machine. We
buffer each tape of the machine with end-marker symbols. If for a
given input string, the head on any output-tape reaches an end-marker,
the machine halts in a non-final state and thus fails to accept the
string. It will be clear that for a string in ${\cal B}_{\infty}$ this
will never happen.

In what follows we denote the tapes of the Turing machine by $t_{i},
i=0,\ldots,N$. $t_{0}$ is the input-tape which is read-only.
Auxiliary tapes necessary to perform computations at Step(i) will
be denoted as $t_{ai},t_{bi}$ etc.
\vskip 0.075in
\noindent Step(1): Determine $n$ for the given input word $\psi$.

We wish to place $n$ $1$'s on $t_{1}$ to determine the depth of the
binary tree (see Remark (1)) where $\psi$ could occur. This is easily
possible using Remark (2). Since the first wedge\footnote{In what
follows we will use the terms ``wedge'' and ``the word representing
the wedge'' interchangeably when there is no danger of confusion.} in
any word of ${\cal B}_{\infty}$ has length $2^{n+2}+7$ for some $n$,
we can use two auxiliary tapes $t_{a1},t_{b1}$ to evaluate $n$.
\vskip 0.075in
\noindent Step(2): Determine the possible strings $s$ on $\{l,r\}$
of length $n$, in lexicographical order.

We now want to obtain the nodes of the tree ${\cal T}_{s}$ (see
Remark (1)) at depth $n$. These nodes will be written on $t_{2}$ in
the lexicographical order from left to right and will be separated
by commas. To achieve this we need two auxiliary tapes $t_{a2},t_{b2}$.
For every $1$ encountered
on $t_{1}$, the tapes $t_{a2}$ and $t_{b2}$ can be used alternately to
generate the nodes of ${\cal T}_{s}$ at successive depths by copying
and prefixing. This is continued till the first blank is encountered on
$t_{1}$, when the process stops and the contents of the last auxiliary
tape written on, are copied to $t_{2}$.
As the formal description of the entire procedure is
cumbersome, we leave it to the reader to check it. It is easy
to see that the limited size of the tapes available, in fact, suffice for
this purpose.
\vskip 0.075in
\noindent Step(3)-(8): Determine $X_{s}, Y_{s}, U_{s}, \bar{U}_{s},
V_{s}, \bar{V}_{s}$ for an $s$ on $t_{2}$

For an $s$ that has been written on tape $t_{2}$,
we wish to determine the corresponding
$X,Y,U,\bar{U},V,\bar{V}$. These will be stored on tapes
$t_{3}$ to $t_{8}$ respectively. As the process is recursive we can use
auxiliary tapes $t_{ai},t_{bi} i=3,\ldots,8$ for each main storage
tape. While the determination of $X,Y$ requires mere copying from
one tape to another, the
determination of $U,\bar{U},V,\bar{V}$ will require the
determination of the parity of the number of $R$'s in $X,Y$.
\vskip 0.075in
\noindent Step(9): Determine the word corresponding to $R_{n}$
in ${\cal B}_{\infty}$.

For each $s$ the wedges that occur in the definition of $R_{n}$
are easily obtained by copying the $X,Y,U,\bar{U},V,
\bar{V}$ in the relevant order to $t_{9}$. Control then returns
to Steps(3)-(8) where $X,Y,U,\bar{U},V,\bar{V}$ corresponding to the next
$s$ are obtained and so forth.

After this the word in $t_9$ is compared symbol-by-symbol with
that in $t_{0}$. If it is the same
the machine halts in a final state or else it halts in a non-final
state.

We observe that if $\psi \in {\cal B}_{\infty}$, then each step in the
process described above, could be carried out on the bounded set of tapes
that was available. Thus if during any of these steps, a head reaches
the end of a particular storage-tape, we are guaranteed that $\psi \not\in
{\cal B}_{\infty}$ and the machine would halt in a non-final state,
rejecting the word. This completes the proof.

\section*{\protect\normalsize\bf 5. Approaching the Onset Of Chaos}

Let us first consider the description of the onset of chaos in the
unimodal case. It might seem at first sight that the description of
this set by means of the language ${\cal A}_{\infty}$ as being rather {\it ad
hoc}. In fact instead of choosing to approach the accumulation point
through a sequence of superstable bifurcations, we might as well have
chosen any of the other sequences available. However, it is easy to
see that the kneading sequences within a periodic window are very
simply related to each other. We could in fact have included every kneading
sequence less than $X_{\infty}$ to form a new language
${\cal C}_{\infty}$. This new language would be described simply as
\begin{displaymath}
{\cal C}_{\infty} = {\cal A}^{<}_{\infty} \bigcup {\cal A}_{\infty}
                                           \bigcup {\cal A}^{>}_{\infty}
\end{displaymath}
where ${\cal A}^{<}_{\infty}$ (${\cal A}^{>}_{\infty}$) denotes the language
containing the kneading sequences to the left (resp. right) of the
superstable sequences in each periodic window.  A proof almost
identical to the one given for ${\cal A}_{\infty}$ however shows that both
${\cal A}^{<}_{\infty}$ and ${\cal A}^{>}_{\infty}$
(and hence ${\cal C}_{\infty}$) are indexed languages.

A more pertinent question is what happens if we are to approach the
onset of chaos from the chaotic side. This question is rather
tricky. For the sake of discussion let us fix the family of maps to be
the logistic family, described by the equation $f(x) = \mu x(1-x)$. It
will be clear that everything would go through for a much larger class
of maps as well.  We could approach the accumulation point $\mu_\infty$
from the right through a sequence of band-merging points. The kneading
sequences at these points are most conveniently described by the so
called $*$-operation, originally discovered by Derrida, Pomeau and
Gervois \cite{derrida}. We give here the definition which best suits
our purpose. Given a word $K''$ on the symbols $\{L,C,R\}$, $R*K''$
is obtained by applying the following transformations simultaneously
on every symbol of $K''$:
\begin{displaymath}
\begin{array}{l}
L \rightarrow RR \\
R \rightarrow RL \\
C \rightarrow RC
\end{array}
\end{displaymath}
We begin with the region of well-developed
chaos at $\mu = 4$. The kneading sequence at this point is
$K''_1 =R(L)^\infty$.  The band-merging points are then obtained through
successive applications of the $*$-operator as
\begin{eqnarray*}
 K''_2 & = & R*K''_1,\\
 K''_3 & = & R*R*K''_1, \\
      & \vdots &      \\
 K''_n & = & (R*)^{n}K''_1, \\
      & \vdots &
\end{eqnarray*}
Now consider the language ${\cal A}''_\infty = \{ K''_n : n=1,2,\ldots\}$. We
could think of this language as another description of the edge of
chaos, this time from the region greater than $\mu_\infty$. In fact
the definition of the $*$-operator immediately confirms that this is
a $0L$ language and hence is also an indexed language (see
\cite{hopcroft}).
Note that we could have discussed the languages
${\cal A}_\infty$ and ${\cal A}''_\infty$ using the $*$-operation as
well (see, for example, \cite{collet}).

We could now ask the same question that we asked before.  What about
other descriptions of the onset of chaos from the chaotic part of the
spectrum? We do not at present have any good answer to this
question. As a preliminary observation we might note that we
could have chosen to approach the onset of chaos through the kneading
sequences $K'_n$ described in section 2.1 and could have
described yet another language
${\cal A}'_\infty = \{ K'_n : n=1,2,3\ldots\}$. Of course this
language is also an indexed language.

However, there is in fact a crucial difference between the language
${\cal A}_{\infty}$ and the languages ${\cal A}'_{\infty}$ or
or ${\cal A}''_{\infty}$, which, though obvious, might be well worth
pointing out: each word in ${\cal A}_\infty$ describes a stable periodic
point, whereas that in ${\cal C}_\infty$ describes an unstable periodic
point. In the chaotic regime the attracting sets would be described by
aperiodic symbol-sequences which are infinitely long. Unfortunately,
classical computation theory does not consider within its domain,
languages whose words might be infinitely long. In fact the behaviour
of classical computational devices on words of infinite length is very
different from their behaviour on words of finite length (see
\cite{eilenberg}). We speculate that a description of computation over
the field of reals, when sufficiently elaborate so as to provide an
analogue of the Chomsky hierarchy might throw light upon these
issues. The recent developments in this direction due to Blum, Shub and
Smale \cite{blum} might provide the seeds of the such a theory.

The drawbacks mentioned above also apply to our discussion of the bimodal
maps. We could as in the unimodal case, choose to approach the
edge from the chaotic side. In the notation of sections 2.2 and
4.2, the language ${\cal B}'_\infty = \{ w(S_n): n=1,2,3\ldots\}$
can be considered as describing the region $D$, thought of as the
boundary of the region $S_\infty$. The proof given in section 4.2
goes through almost unaltered even for this case. Thus
${\cal B}'_\infty$ is a context-sensitive language.

\section*{\protect\normalsize\bf 6. On complex descriptions.}

In conclusion we would like to place our results in perspective.
What paradigm do these results suggest for a definition of
complexity? In order to address this question we must first inquire,
into the process of description. The scientific description of phenomena
normally involves two aspects. The first is the specification of
the {\em model-class}. The second aspect
involves the description of the phenomenon at hand, with respect
to this model-class. Let us call this process, {\em interpretation}.
For a description to be ``useful'' we must ensure that that both the
model-class and the interpretation have been specified in finite terms.
Given this rather simplistic picture of the modeling process we now
ask, how complexity arises or, more specifically, why are some phenomena
more complex than others? Consider a phenomenon which resists finite
interpretation with respect to a certain model class. In order to
describe it we would, then, have to construct a ``larger'' model-class.
This could be regarded as signalling complexity.

To illustrate this in the context of our results, consider the
automata or equivalently the grammars in the Chomsky hierarchy as
representing model-classes. Let the behaviour of maps (unimodal or
bimodal) represent the entire class of phenomena to be described.
The symbolic dynamics, giving rise to a language, and the explicit
construction of a grammar (corresponding to a given model-class)
generating that language, at each value in the
parameter space, can serve as an interpretation for the phenomenon.
Together these constitute a description of the behaviour of the map in
question. Now, at the onset of chaos we observe that we are forced to
change our model-class (in the unimodal case, for example, from the
regular grammars to the indexed grammars). In fact in our case we have
proved, that no finite interpretation can be obtained of the edge of
chaos in terms of the older model-class. This describes the complexity
at the onset of chaos.

Finally we would like to point out that unlike conventional
statistical-mechanics that relies on numerical or quantitative
classes (say using critical exponents), the paradigm suggested above
favours a more qualitative definition of complexity. What one is
tempted to observe, is that, complexity is not so much a matter
of number, as it is of mechanism.

\vskip 0.5in

\noindent {\bf Acknowledgement}: I would like to acknowledge Spenta
R. Wadia for having introduced me to the subject of complex systems and
for his constant guidance during the course of my work. I thank
Gautam Mandal, Avinash Dhar and Spenta R. Wadia for their
encouragement and support at all stages. I am grateful for the
endless useful discussions that I have  had with them which have
benefited me enormously. I also wish to thank Jim Crutchfield for
having communicated to me details regarding his work and for pointing
out an error in the manuscript.

\vskip 0.5in
\noindent {\bf Note added}: After most of this work was completed, we
discovered that Crutchfield and Young \cite{crutchfield2,crutchfield3}
have also shown that the formal language corresponding to the unimodal
case is an indexed language. Their approach, based on the
$\varepsilon$-machine reconstruction, is more general but different
from the one (based on symbolic dynamics) pursued in this paper.


\begin{thebibliography}{99}
\bibitem{wolf} S. Wolfram, ``Universality and complexity in cellular
 automata'', {\em Physica D}, {\bf 10}, (1984) 1.
\bibitem{chaitin} G. Chaitin, {\em Algorithmic Information Theory},
Cambridge University Press (1987).
\bibitem{avinash} A. Dhar, G. Mandal, P. Lakdawala, S. R. Wadia,``The
Role of Initial Conditions in the Classification of
the Rule-space of Cellular Automata Dynamics'', {\em Phys. Rev. E},
{\bf 51} (1995) 3032.
\bibitem{crutchfield} J. Crutchfield, ``The calculi of emergence: computation,
dynamics and induction'', {\em Physica D}, {\bf 75}, (1994) 11.
\bibitem{collet} P. Collet, J.P. Eckmann, {\em Iterated Maps of the Interval
as Dynamical Systems}, Birkhauser (1980).
\bibitem{mackay} R.S. Mackay, C. Tresser, ``Boundary of topological chaos
for bimodal maps of the interval'', {\em J. London Math. Soc.}, {\bf 37},
(1988) 164.
\bibitem{milnor} J. Milnor, W. Thurston, ``On iterated maps of the interval'',
in {\em Lecture Notes in Math. 1342}, Springer-Verlag (1988).
\bibitem{feigenbaum} M.J. Feigenbaum, ``Universal behaviour in nonlinear
systems'', {\em Physica D}, {\bf 7}, (1983) 16.
\bibitem{hopcroft} J.E. Hopcroft, J.D. Ullman, {\em Introduction to
Automata Theory, Languages and Computation}, Addison-Wesley, USA (1979).
\bibitem{ogden} W. Ogden, ``Intercalation theorems for stack languages'',
in {\em Proc. First Annual ACM Symposium on the Theory of Computing}, (1969)
 31.
\bibitem{hayashi} T. Hayashi, ``On Derivation Trees of Indexed Grammars
-- An Extension of the uvwxy-Theorem --'',{\em Publ. RIMS, Kyoto Univ.},
{\bf 9} (1973) 61.
\bibitem{derrida} B. Derrida, A. Gervois, Y. Pomeau,``Iterations of
endomorphisms on the real axis and representations of numbers'',{\em
Ann. de l'Inst. Henri Poincare}, {\bf 29} (1978) 305.
\bibitem{eilenberg} S. Eilenberg,{\em Automata, Languages and Machines},
vol. A, Academic Press, New York (1974).
\bibitem{blum} L. Blum, M. Shub, S. Smale,``On a Theory of Computation
and Complextiy over the Real Numbers: {\it NP}-Completeness, Recursive
Functions and Universal Machines'',{\em Bull. Am. Math. Soc.},{\bf 21}
(1989) 1.
\bibitem{crutchfield2}  J. Crutchfield, K. Young, `` Inferring Statistical
Complexity'', {\em Phys. Rev. Lett.},{\bf 63} (1989) 105.
\bibitem{crutchfield3} J. Crutchfield, K. Young, `` Computation at
the Onset Of Chaos'', in {\em Complexity, Entropy and the Physics of
Information}, Addison-Wesley (1990).
\end{thebibliography}
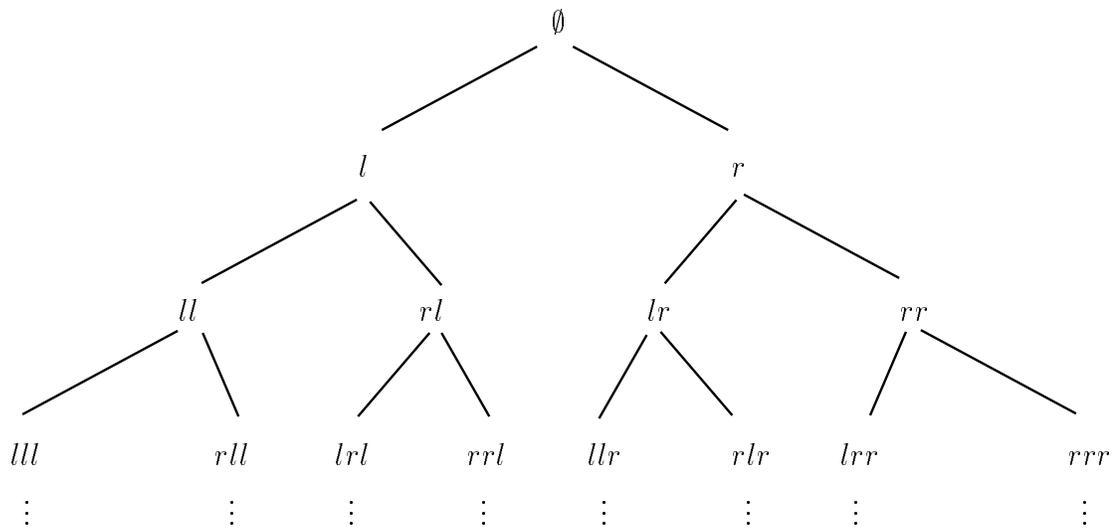
\end{document}